\documentclass[english,aps,prc,nofootinbib,superscriptaddress,twocolumn,letter]{revtex4}
\usepackage[latin9]{inputenc}
\usepackage{hyperref}
\hypersetup{colorlinks=true, linkcolor=blue, citecolor=blue, urlcolor=blue}

\usepackage{breakurl}
\usepackage{graphicx}
\usepackage{epsf}
\usepackage{epsfig}
\usepackage{amssymb,amsmath}
\usepackage[usenames]{color}
\usepackage{amssymb}
\usepackage{times}
\usepackage{comment}
\usepackage{setspace}

\usepackage[normalem]{ulem} 
\usepackage{float} 

\usepackage{graphicx}

\allowdisplaybreaks



\usepackage{hyperref}

\begin{document}

\preprint{This line only printed with preprint option}

\title{Thermodynamics of Heavy Quarkonium in a Bayesian Holographic QCD model}


\author{Liqiang Zhu}
\email{zhuliqiang@mails.ccnu.edu.cn}
\affiliation{Key Laboratory of Quark and Lepton Physics (MOE) and Institute of Particle Physics, Central China Normal University, Wuhan 430079, China}

\author{Ou-Yang Luo}
\email{luoouyangedu@qq.com}
\affiliation{School of Nuclear Science and Technology, University of South China, Hengyang 421001, China}

\author{Xun Chen}
\email{chenxun@usc.edu.cn}
\affiliation{Key Laboratory of Quark and Lepton Physics (MOE) and Institute of Particle Physics, Central China Normal University, Wuhan 430079, China}
\affiliation{School of Nuclear Science and Technology, University of South China, Hengyang 421001, China}
\affiliation{Key Laboratory of Advanced Nuclear Energy Design and Safety, Ministry of Education, Hengyang, 421001, China}
\affiliation{INFN -- Istituto Nazionale di Fisica Nucleare -- Sezione di Bari, Via Orabona 4, 70125 Bari, Italy}

\author{Kai Zhou}
\email{zhoukai@cuhk.edu.cn}
\affiliation{School of Science and Engineering, The Chinese University of Hong Kong, Shenzhen (CUHK-Shenzhen), Guangdong, 518172, China}

\author{Hanzhong Zhang}
\email{zhanghz@mail.ccnu.edu.cn}
\affiliation{Key Laboratory of Quark and Lepton Physics (MOE) and Institute of Particle Physics, Central China Normal University, Wuhan 430079, China}

\author{Defu Hou}
\email{houdf@mail.ccnu.edu.cn}
\affiliation{Key Laboratory of Quark and Lepton Physics (MOE) and Institute of Particle Physics, Central China Normal University, Wuhan 430079, China}

\begin{abstract}

Leveraging high-precision lattice QCD data on the equation of state and baryon number susceptibility at vanishing chemical potential, we construct a Bayesian holographic QCD model and systematically analyze the thermodynamic properties of heavy quarkonium in QCD matter under varying temperatures and chemical potentials.
We compute the quark-antiquark interquark distance, potential energy, entropy, binding energy, and internal energy. 
We present detailed posterior distribution results of the thermodynamic quantities of heavy quarkonium, including maximum a posteriori (MAP) value estimates and 95\% confidence levels (CL). Through numerical simulations and theoretical analysis, we find  that increasing temperature and chemical potential decrease the quark distance, thereby facilitating the dissociation of heavy quarkonium and leading to suppressed potential energy. The increase in temperature and chemical potential also raise the entropy and entropy force, further accelerating the dissociation of heavy quarkonium. The calculated results of binding energy indicate that higher temperature and chemical potential enhance the tendency of heavy quarkonium to dissociate into free quarks. Internal energy also increases with rising temperature and chemical potential. These findings provide significant theoretical insights into the properties of strongly interacting matter under extreme conditions and lay a solid foundation for the interpretation and validation of future experimental data. Finally, we also present the results for the free energy, entropy, and internal energy of single quark.

\end{abstract}
\maketitle
\date{\today}

\section{Introduction}
\label{sec1}
Ultra-relativistic heavy-ion collisions at the Relativistic Heavy Ion Collider (RHIC) and the Large Hadron Collider (LHC) are believed to have created a new state of matter known as the Quark-Gluon Plasma (QGP)\cite{PHENIX:2004vcz, STAR:2010vob,Chen:2024aom,Wang:2022fwq,Wang:2021xpv,Wang:2019vhg,Xu:2018fog,Jin:2018lbk,Liu:2017rjm,Zhu:2025edv,Li:2024uzk,Xie:2024xbn,He:2023zin,Chen:2024eaq,Xu:2016fns,Ma:2023zfj}.
In these extreme environments, heavy quark systems serve as sensitive probes, enabling the investigation of QGP properties and the underlying strong interaction dynamics governed by Quantum Chromodynamics (QCD)\cite{Bulava:2019iut, Li:2023ciq, Yang:2015aia, Zhou:2020ssi,Shi:2021qri,Zhou:2014hwa,Zhou:2014kka}. 
Heavy quark-antiquark pairs form bound states through strong interactions mediated by gluons. It's pointed out that due to the formation of the hot and dense QGP medium\cite{Matsui:1986dk}, these pairs have a maximum dissociation distance beyond which the pair becomes unstable and separates.
The interquark potential governs the formation and stability of these bound states.
As temperature and density intensify, the medium's color screening effect weakens the binding potential between heavy quarks, prompting dissociation -- a hallmark phenomenon signifying the transition from confinement to deconfinement.
Hence, exploring the in-medium heavy quark potential has emerged as a pivotal endeavor for deepening our undertanding of hadronic structure and shedding light on the fundamental nature of QCD matter.
Furthermore, the concept of the holographic potential for heavy quark-antiquark pairs was systematically introduced for the first time in \cite{Maldacena:1998im}, providing a theoretical foundation for investigating the structure of hadrons and the dynamics of QCD \cite{Zhu:2024dwx,Zhu:2024uwu,Yang:2015aia, Zhou:2020ssi, Andreev:2006ct, Andreev:2006eh, Andreev:2006nw, He:2010bx, Colangelo:2010pe, DeWolfe:2010he, Li:2011hp, Fadafan:2011gm, Fadafan:2012qy, Cai:2012xh, Li:2012ay, Fang:2015ytf, Zhang:2015faa, Ewerz:2016zsx, Chen:2017lsf, Arefeva:2018hyo, Chen:2018vty, Bohra:2019ebj, Chen:2019rez, Zhou:2021sdy, Chen:2020ath, Chen:2021gop}.

Heavy-ion collisions thus provide a unique and powerful experimental setting for probing QCD matter under extreme conditions. However, fully interpreting these experiments through lattice QCD simulations at finite chemical potentials is hindered by the fermion sign problem\cite{Fodor:2001au, Muroya:2003qs}, which severely limits numerical calculations in regions of high density.
Although several innovative methodologies \cite{Lang:1982tj, Hoek:1987uy, Michael:1990az, Takahashi:2002bw, Aoki:2005vt, Ratti:2005jh, Bicudo:2007xp, Luscher:2010iy, Hasenfratz:1983ba, Jiang:2022zbt, Yu:2023hzl,Zhou:2023pti} have been developed to circumvent this obstacle, a complete and rigorous first-principles theoretical description of strongly coupled QCD matter under experimental conditions remains elusive. Consequently, complementary theoretical frameworks, particularly holographic QCD approaches inspired by gauge/gravity duality from string theory, have gained prominence.
Presently, the "top-down" approach is centered on constructing a realistic holographic QCD theory derived from string theory \cite{He:2007juu, Burrington:2004id, Sakai:2005yt, Sakai:2004cn, Erdmenger:2020lvq}, whereas the "bottom-up" approach prioritizes the development of holographic model \cite{Braga:2017fsb, Braga:2018fyc, Ferreira:2019nkz, He:2010ye, Jokela:2024xgz, Arefeva:2023fky, Rougemont:2023gfz} informed by experimental observations and lattice QCD data. For instance, the incorporation of black hole configurations in five-dimensional space-time to elucidate boundary phenomena at finite temperatures, along with the exploration of broader theoretical frameworks, constitutes pivotal research directions in this field \cite{Fadafan:2012qy, Zhang:2015faa, Rey:1998bq, Brandhuber:1998bs, Noronha:2010hb, Zhang:2016jns, Chen:2021bkc, Guo:2023zjx}.

Within the framework of holographic theory, the dissociation of heavy quark bound states in QGP medium have been extensively studied \cite{Iatrakis:2015sua, Jena:2022nzw}. 
These heavy quark bound states can be geometrically represented by open strings \cite{Song:2007gm, Escobedo:2013tca} whose endpoints correspond to a heavy quark and antiquark pair located at the spacetime boundary, separated by an interquark distance $L$. 
In such models, the thermodynamics of QCD matter at finite temperature is mapped onto a five-dimensional black hole geometry, where the gravitational attraction of the black hole horizon dynamically captures the medium-induced effects experienced by the heavy quark-antiquark string.
The string serves as a geometric representation of a bound state in QCD within the holographic framework. Initially, the string remains static at the boundary. However, influenced by the gravitational effects of the background spacetime metric, the string is gradually drawn towards the black hole's event horizon. This dynamic evolution of the string captures the microscopic mechanisms underlying quark pair dissociation. As the string approaches the horizon, the system achieves equilibrium, and the binding energy of the string diminishes to a critical value, resulting in the dissociation of the quark-antiquark pair. The timescale associated with this process known as dissociation time of the quark-antiquark pair, as a critical parameter in the study of quarkonium suppression, provides a key indicator of the non-equilibrium properties of strongly interacting media under extreme conditions. This timescale provides essential theoretical insights into the formation and evolution of the QGP. Moreover, examining the variations in dissociation time not only helps elucidate the strongly coupled dynamics of QCD but also explains the experimentally observed suppression of heavy quarkonium states. Such analyses significantly enhance our understanding of matter under extreme conditions in high-energy nuclear physics.

Based on the holographic QCD models described in prior works \cite{Chen:2024ckb, Chen:2024mmd, Zhu:2025gxo}, the main research objective of this paper is to analyze the impact of varying temperature and chemical potential on the thermodynamic properties of heavy quarkonium in a 2+1 flavor system. The structure of this paper is organized as follows: Section \ref{sec2} provides a brief review of the holographic model that incorporates information about the QCD phase transition.  In Section \ref{sec3}, we analyze the dissociation distance, potential energy, entropy, entropy force, binding energy, and internal energy of the heavy quark-antiquark pair under different temperature and chemical potential conditions. Section \ref{sec4} examines the effects of temperature and chemical potential on the thermodynamic properties of a single quark. Finally, Section \ref{sec5} summarizes the research findings and presents concluding remarks.

\section{The Setup}
\label{sec2}

This section will provide an overview of the Einstein-Maxwell-dilaton (EMD) model. Within the context of the string frame, the action for the EMD model reads \cite{Yang:2015aia, Zhou:2020ssi, Chen:2018vty, Chen:2019rez, Chen:2020ath, Chen:2024ckb, Chen:2024mmd, He:2013qq, Yang:2014bqa, Dudal:2017max, Dudal:2018ztm}
\begin{align}
\begin{split}
S_b&=\frac{1}{16 \pi G_5} \int d^5 x \sqrt{-g^s} e^{-2 \phi_s}\\
&\quad\times\left[R_s-\frac{f_s\left(\phi_s\right)}{4} F^2+4 \partial_\mu \phi_s \partial^\mu \phi_s-V_s\left(\phi_s\right)\right].
\label{Eq:actionsb}
\end{split}
\end{align}
Here, $f(\phi)$  is the gauge kinetic function that is coupled to the Maxwell field $A_{\mu}$, while $V(\phi)$ defines the dilaton field's potential. The parameter $G_{5}$ corresponds to the Newton constant in five-dimensional spacetime. By solving the equations of motion (EoMs), the functional forms of $f(\phi)$ and $V(\phi)$ can be consistently determined. The action is rewritten in the Einstein frame from the string frame through a specific set of transformations.
\begin{align}
\begin{split}
\phi_s=\sqrt{\frac{3}{8}} \phi, \quad &g_{\mu \nu}^s=g_{\mu \nu} e^{\sqrt{\frac{2}{3}} \phi},\quad
f_s\left(\phi_s\right)=f(\phi) e^{\sqrt{\frac{2}{3}} \phi},\\
&V_s\left(\phi_s\right)=e^{-\sqrt{\frac{2}{3}} \phi} V(\phi).
\end{split}
\end{align}
In the Einstein frame, the action is expressed as follows:
\begin{align}
\begin{split}
S_b&=\frac{1}{16 \pi G_5} \int d^5 x \sqrt{-g}\\
&\quad\times\left[R-\frac{f(\phi)}{4} F^2-\frac{1}{2} \partial_\mu \phi \partial^\mu \phi-V(\phi)\right].
\end{split}
\end{align}
We propose the following metric ansatz:
\begin{equation}
d s^2=\frac{L^2 e^{2 A(z)}}{z^2}\left[-g(z) d t^2+\frac{d z^2}{g(z)}+d \vec{x}^2\right],
\end{equation}
In this context, $z$ serves as the holographic coordinate in the fifth dimension, and the $\mathrm{AdS}_{5}$ space is characterized by a fixed radial parameter $R_{\mathrm{AdS}}=1$. By employing the aforementioned metric ansatz, one can derive the corresponding equations of motion and constraints for the background fields
\begin{align}
\begin{split}
\phi^{\prime \prime}&+\phi^{\prime}\left(-\frac{3}{z}+\frac{g^{\prime}}{g}+3 A^{\prime}\right)-\frac{e^{2 A}}{z^2 g} \frac{\partial V}{\partial \phi}\\
&+\frac{z^2 e^{-2 A} A_t^{\prime 2}}{2g} \frac{\partial f}{\partial \phi}=0,
\end{split}
\end{align}
\begin{equation}
A_t^{\prime \prime}+A_t^{\prime}\left(-\frac{1}{z}+\frac{f^{\prime}}{f}+A^{\prime}\right)=0,
\end{equation}
\begin{equation}
g^{\prime \prime}+g^{\prime}\left(-\frac{3}{z}+3 A^{\prime}\right)-\frac{e^{-2 A} A_t^{\prime 2} z^2 f}{L^2}=0,
\end{equation}
\begin{align}
\begin{split}
A^{\prime \prime}&+\frac{g^{\prime \prime}}{6 g}+A^{\prime}\left(-\frac{6}{z}+\frac{3 g^{\prime}}{2 g}\right)-\frac{1}{z}\left(-\frac{4}{z}+\frac{3 g^{\prime}}{2 g}\right)\\
&+3 A^{\prime 2} +\frac{e^{2 A} V}{3 z^2 g}=0,
\end{split}
\end{align}
\begin{equation}
A^{\prime \prime}-A^{\prime}\left(-\frac{2}{z}+A^{\prime}\right)+\frac{\phi^{\prime 2}}{6}=0.
\end{equation} 
Among these five equations, only four are independent in a linear sense. The infrared (IR) boundary conditions for the equations of motion (EoMs) near the horizon ($z=z_{h}$) represent the boundary of the black hole, which is physically associated with the concept of temperature. This can be expressed as follows:
\begin{equation}
A_t\left(z_h\right)=g\left(z_h\right)=0.
\end{equation}
As the IR boundary is approached $z \rightarrow z_{h}$, we impose the condition that the metric in the string frame converges to $\mathrm{AdS}_{5}$. The conditions at the ultraviolet (UV) boundary ($z=0$) are as follows:
\begin{equation}
A(0)=-\sqrt{\frac{1}{6}} \phi(0), \quad g(0)=1, \quad A_t(0)=\mu+\rho^{\prime} z^2+\cdots,
\end{equation}
wehre $\mu$ denotes the baryon chemical potential, while $\rho^{\prime}$ scales proportionally with the baryon number density, The baryon number density can be calculated based on \cite{Critelli:2017oub, Zhang:2022uin}
\begin{equation}
\begin{aligned}
\rho & =\left|\lim _{z \rightarrow 0} \frac{\partial \mathcal{L}}{\partial\left(\partial_z A_t\right)}\right| \\
& =-\frac{1}{16\pi G_5} \lim _{z \rightarrow 0}\left[\frac{\mathrm{e}^{A(z)}}{z} f(\phi) \frac{\mathrm{d}}{\mathrm{d} z} A_t(z)\right].
\end{aligned}
\end{equation}
$\mu$ is related to the quark-number chemical potential $\mu=3\mu_{q}$. 
This allows us to derive
\begin{align}
\begin{split}
g(z)&=1-\frac{1}{\int_{0}^{z_{h}}dxx^{3}e^{-3A(x)}}\\
&\quad\times\Bigg[\int_{0}^{z}dxx^{3}e^{-3A(x)}+\frac{2c\mu^{2}e^{k}}{(1-e^{-cz_{h}^{2}})^{2}}\mathrm{det}\mathcal{G}\Bigg],
\end{split}
\end{align}
\begin{align}
\phi^{\prime}(z)=\sqrt{6\left(A^{\prime 2}-A^{\prime \prime}-\frac{2}{z} A^{\prime}\right)},
\end{align}
\begin{align}
A_t(z)=\mu\frac{e^{-cz^{2}}-e^{-cz_{h}^{2}}}{1-e^{-cz_{h}^{2}}},
\end{align}
\begin{align}
\begin{split}
V(z)&=-3z^{2}ge^{-2A}\Bigg[A^{\prime\prime}+A^{\prime}\bigg(3A^{\prime}-\frac{6}{z}+\frac{3g^{\prime}}{2g}\bigg)\\
&\quad-\frac{1}{z}\bigg(-\frac{4}{z}+\frac{3g^{\prime}}{2g}\bigg)+\frac{g^{\prime\prime}}{6g}\Bigg],
\end{split}
\end{align}
where
\begin{align}
\operatorname{det}\mathcal{G}=\left|\begin{array}{ll}
\int_0^{z_h} d y y^3 e^{-3 A(y)} & \int_0^{z_h} d y y^3 e^{-3 A(y)-c y^2} \\
\int_{z_h}^z d y y^3 e^{-3 A(y)} & \int_{z_h}^z d y y^3 e^{-3 A(y)-c y^2}
\end{array}\right|.
\end{align}
The Hawking temperature \cite{Hawking:1982dh,Natsuume:2014sfa} is defined by
\begin{small}
\begin{align}
\begin{split}
T & =\frac{\left|g^{\prime}(z)\right|}{4\pi}\\
& =\frac{z_h^3 e^{-3 A\left(z_h\right)}}{4 \pi \int_0^{z_h} d y y^3 e^{-3 A(y)}}\Bigg[1+ \\
&\frac{2 c \mu^2 e^k\left(e^{-c z_h^2} \int_0^{z_h} d y y^3 e^{-3 A(y)}-\int_0^{z_h} d y y^3 e^{-3 A(y)} e^{-c y^2}\right)}{(1-e^{-c z_h^2})^2} \Bigg].
\end{split}
\end{align}   
\end{small}
In order to achieve an analytical solution, we make use of
\begin{align}
A(z)= d\ln(a z^2 + 1) + d\ln(b z^4 + 1).
\end{align}
The gauge kinetic function $f(z)$ is defined as:
\begin{align}
f(z)=e^{cz^{2}-A(z)+k}.
\end{align}
In the string frame, the potential of a quark-antiquark pair is determined via $A_{s}(z)=A(z)+\sqrt{\frac{1}{6}}\phi(z)$. Standard techniques enable the computation of their separation distance and interaction potential \cite{Colangelo:2010pe, Li:2011hp, Chen:2017lsf, Rey:1998bq, Guo:2023zjx}. The dynamics of the string world-sheet are described by the Nambu-Goto action in the form shown below
\begin{align}
S_{NG}=-\frac{1}{2\pi\alpha^{\prime}}\int d^{2}\xi\sqrt{-\mathrm{det}g_{ab}}.
\end{align}
In this context, $g_{ab}$ refers to the induced metric, defined as:
\begin{align}
g_{ab}=g^{s}_{MN}\partial_{a}X^{M}\partial_{b}X^{N},\quad a,b=0,1,
\end{align}
and $\alpha^{\prime}$ is associated with the string tension and is assigned a value of 1. In this context, $X^{M}$ represents the coordinates, while $g^{s}_{MN}$ denotes the metric in the string frame. For the calculation of the heavy quark-antiquark potential, the string is anchored at a static quark-antiquark pair located at $x_{1}=-L/2$ and $x_{1}=L/2$. The most straightforward parametrization of the string world-sheet coordinates is $\xi^{0}=t$ and $\xi^{1}=x_{1}$. In this scenario, the effective Nambu-Goto action can be represented as:
\begin{align}
S_{NG}=-\frac{1}{2\pi T}\int^{L/2}_{-L/2}dx_{1}\sqrt{k_{1}(z)\frac{dz^{2}}{dx_{1}^{2}}+k_{2}(z)}
\end{align}
where
\begin{align}
\begin{split}
k_{1}&=\frac{e^{4}A_{s}}{z^{2}},\\
k_{2}&=\frac{e^{4}A_{s}}{z^{2}}g(z).
\end{split}
\end{align}
Based on the research in \cite{Andreev:2006nw,Rey:1998bq,Witten:1998zw,Brandhuber:1998er,Gross:1998gk,Giataganas:2011nz}, the Wigner-Wilson loop's expectation value is linked to the on-shell string action by:
\begin{align}
\langle W(\mathcal{C})\rangle =\int DXe^{-S_{NG}}\simeq e^{-S_{\mathrm{on-shell}}}.
\end{align}
Here, $\mathcal{C}$ represents a closed loop in spacetime. The heavy-quark potential is defined as \cite{Maldacena:1998im,Ewerz:2016zsx,Rey:1998ik}:
\begin{align}
\langle W(\mathcal{C})\rangle =\sim e^{-V(r,T)/T}
\end{align}
$L$ denotes the distance between the quark-antiquark pair. Consequently, calculating the potential requires solving the on-shell string world-sheet action. Based on the standard framework \cite{Yang:2015aia,Andreev:2006nw,Colangelo:2010pe,Li:2011hp,Chen:2017lsf,Chen:2020ath}, we establish an effective ``Hamiltonian".
\begin{align}
\mathcal{H}=z^{\prime}\frac{\partial\mathcal{L}}{\partial z^{\prime}}-\mathcal{L}=\frac{k_{2}(z)}{\sqrt{k_{1}(z)z^{\prime 2}+k_{2}(z)}}
\end{align}
Here, $z^{\prime}=\frac{dz}{dx_{1}}$, and solving for $z^{\prime}$ forms the equation
\begin{align}
\frac{k_{2}(z)}{\sqrt{k_{1}(z)z^{\prime 2}+k_{2}(z)}}=\frac{k_{2}(z_{0})}{\sqrt{k_{2}(z_{0})}}
\end{align}
Here, $z_{0}$ marks the vertex where the quark-antiquark string connects,  with values from $z_{0}=0$ to $z_{0}=z_{h}$. This allows us to calculate the interquark distance and the free energy,
\begin{align}
\begin{split}
L&=\int_{-\frac{L}{2}}^{\frac{L}{2}}dx=2\int_{0}^{z_{0}}dz\frac{1}{z^{\prime}}\\
&=2\int_{0}^{z_{0}}\bigg[\frac{k_{2}(z)}{k_{1}(z)}\bigg(\frac{k_{2}(z)}{k_{2}(z_{0}}-1\bigg)\bigg]^{-1/2}
\end{split}
\end{align}
\begin{small}
\begin{align}
\begin{split}
V_{Q\Bar{Q}}&=\frac{1}{\pi}\Bigg(\int_{0}^{z_{0}}dz\Bigg(\sqrt{\frac{k_{2}(z)-k_{1}(z)}{k_{2}(z)-k_{2}(z_{0})}}-\bigg(\frac{1}{z^{2}}+\frac{2\sqrt{-6ad}}{z}\bigg)\Bigg)\\
&\quad -\bigg(\frac{1}{z_{0}}-2\sqrt{-6ad}\mathrm{ln}(z_{0})\bigg)\Bigg)
\end{split}
\end{align}
\end{small}
In our computational framework, the potential energy of the heavy quarkonium is equal to its free energy ($F_{Q\Bar{Q}}=V_{Q\Bar{Q}}$).
The potential is regularized by subtracting the UV divergent term.

The entropy of the quark-antiquark pair is derived as:
\begin{align}\label{eq30}
S_{Q\Bar{Q}}=-\frac{\partial F_{Q\Bar{Q}}}{\partial T}=-\frac{\partial F_{Q\Bar{Q}}}{\partial z_{h}}\frac{\partial z_{h}}{\partial T},
\end{align}
where $T$ represents the QGP temperature, and the binding energy can be given by\cite{Ewerz:2016zsx}
\begin{align}\label{eq31}
E_{Q\Bar{Q}}=F_{Q\Bar{Q}}-2F_{Q},
\end{align}
$F_{Q}$ represents the free energy of a single quark, which can be computed as shown in Ref. \cite{Zhou:2020ssi}, with the integral's upper limit fixed at $z_{h}$. Specifically:
\begin{small}
\begin{align}\label{eq32}
\begin{split}
\frac{F_{Q}}{\sqrt{\lambda}}&=\frac{1}{2\pi}\Bigg(\int_{0}^{z_{h}}dz\bigg(\sqrt{k_{1}(z)}-\bigg(\frac{1}{z^{2}}+\frac{2\sqrt{-6ad}}{z}\bigg)\Bigg)\\
&\quad -\bigg(\frac{1}{z_{h}}-2\sqrt{-6ad}\mathrm{ln}(z_{h})\bigg)\Bigg).
\end{split}
\end{align}
\end{small}
In this paper, we set $\sqrt{\lambda} =1$ for convenience. The internal energy of the quark-antiquark pair, as described in Refs. \cite{Bali:2011qj,Ewerz:2016zsx}, is
\begin{align}\label{eq33}
U_{Q\Bar{Q}}=F_{Q\Bar{Q}}+TS_{Q\Bar{Q}}+\mu N_{Q\Bar{Q}},
\end{align}
with $N_{Q\Bar{Q}}$ related to the baryon number density.

For the 2+1 flavor system, the six parameters of our model were determined using Bayesian inference based on the equation of state (EoS) and baryon number susceptibility data from lattice QCD. We employed the posterior model parameters obtained through Bayesian inference, including the 95\% confidence levels (CL) and the maximum a posteriori (MAP) values \cite{Zhu:2025gxo}, As indicated in Table \ref{table:parameter}.

\begin{table}[htbp] 
\centering 
\begin{tabular}{|c|c|c|c|} 
  \hline
  \multicolumn{4}{|c|}{Posterior  95\% CL} \\ \hline
  Parameter & min & max & MAP \\ \hline
  $a$ & 0.229 & 0.282 & 0.252 \\ \hline
  $b$ & 0.019 & 0.027 & 0.023 \\ \hline
  $c$ & -0.261 & -0.231 & -0.245 \\ \hline
  $d$ & -0.143 & -0.127 & -0.135 \\ \hline
  $k$ & -0.871 & -0.808 & -0.843 \\ \hline
  $G_{5}$ & 0.388 & 0.406 & 0.397 \\ \hline
\end{tabular}
\caption{In the 2+1 flavor system, the 95\% CL ranges and MAP values for the six model parameters $a\;$,$b\;$,$c\;$,$d\;$,$k\;$ and $G_{5}$ are determined through Bayesian inference \cite{Zhu:2025gxo}.} 
\label{table:parameter}
\end{table}

\section{THERMODYNAMICS OF HEAVY QUARKONIUM}
\label{sec3}

\begin{figure*}[tbp!]
    \centering
    \includegraphics[width=0.95\textwidth]{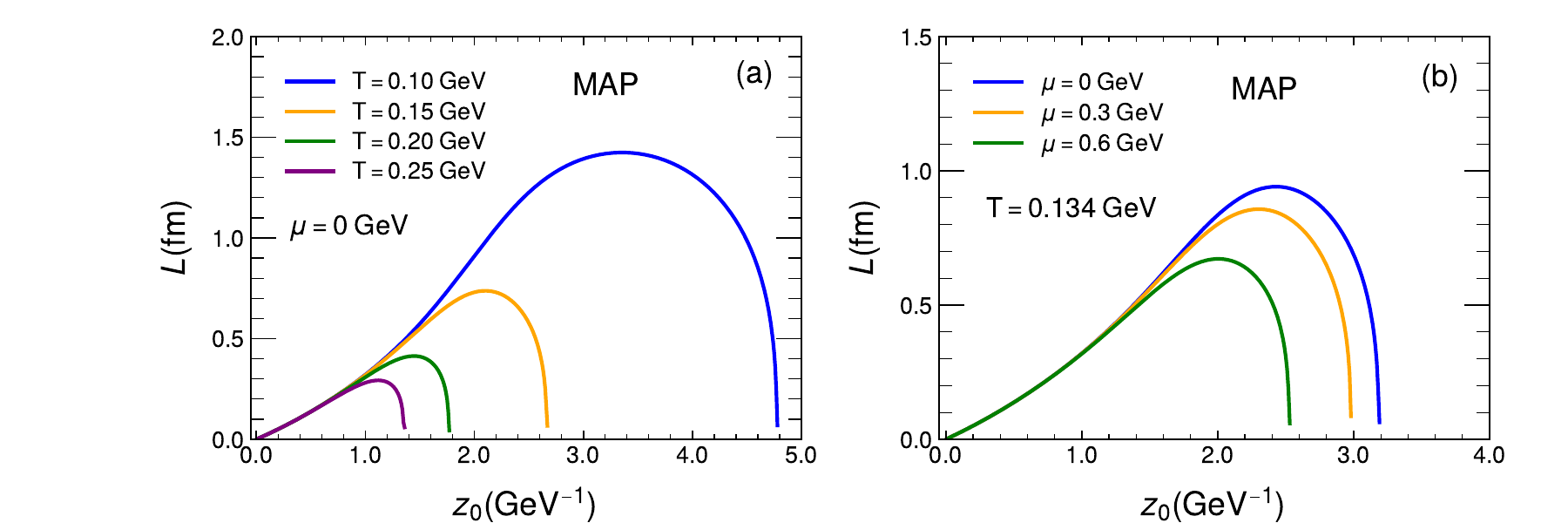}
    \caption{When selecting MAP setup for the Bayesian Holographic model, the calculated interquark distance $L$ as a function of $z_{0}$ for the quark-antiquark pair at different temperatures when $\mu=0$ (a) and at different chemical potential when $T=0.134\;\mathrm{GeV}$ (b).
}
\label{z0-L}
\end{figure*}

In this section, we systematically investigate the thermodynamic properties of heavy quark-antiquark pairs in a 2+1 flavor system, focusing on the effects of temperature and chemical potential on these properties to elucidate their behavior within a strongly interacting medium and the mechanism underlying their transition from the confined phase to the deconfined phase. 
Specifically, this chapter is divided into five subsections:  (A) The interquark distance of heavy quark-antiquark pairs, in which we explore how the dissociation distance varies with temperature and chemical potential, as well as its relationship with confinement effects; (B) The potential energy of heavy quark-antiquark pairs, where we analyze the composition of the potential energy and the phenomenon of its truncation under the influence of temperature and chemical potential; (C) The entropy and entropy force of heavy quark-antiquark pairs, studying the variation of entropy with temperature and chemical potential to reveal the degree of disorder within the system; (D) The binding energy of heavy quark-antiquark pairs, investigating how the binding energy changes with temperature and chemical potential and its connection to quark dissociation; (E) The internal energy of heavy quark-antiquark pairs, analyzing the variation of internal energy with temperature and chemical potential and its role in thermodynamics.
Through these studies, we aim to achieve a comprehensive understanding of the thermodynamic behavior of quark-antiquark pairs in a 2+1 flavor system.

\subsection{The dissociation of heavy quark-antiquark pairs}

In this subsection, we investigate the behavior of the interquark distance $L$ of heavy quark-antiquark pairs in a 2+1 flavor system as a function of the $z_{0}$. Our calculated results of $L(z_0)$ with the maximum a posterior (MAP) inference setup for the Bayesian holographic model\cite{Zhu:2025gxo} is depicted in Fig. \ref{z0-L}, with Fig. \ref{z0-L} (a) shows the results of calculating the effect of different temperatures on $L$ with the chemical potential fixed at zero, and Fig. \ref{z0-L} (b) shows the results of calculating the effect of the chemical potential on $L$ with the temperature fixed at $T=0.134\;\mathrm{GeV}$.

As illustrated in Fig. \ref{z0-L}, it has been observed that the interquark distance $L$ increases with $z_{0}$ until it reaches a peak value known as the dissociation distance $L_{\mathrm{max}}$. After this maximum point, further increases in $z_{0}$ lead to a decrease in the interquark distance $L$. This trend suggests that the quark-antiquark pairs transition into a deconfined state, resulting in the melting of the string that connects them. Consequently, the bound state of the quark-antiquark pairs dissociates, and they eventually become free heavy quarks. This phenomenon highlights the color screening effect of the medium on these bound states, which means that partons in the medium (QGP) can ``shield'' the interactions between the quark and antiquark pairs, preventing their tight binding.

By analyzing the behavior of the dissociation distance, we find that as the temperature and chemical potential increase, the dissociation distance $L_{\mathrm{max}}$ gradually becomes smaller. This is related to the increasing parton density in the medium with increasing temperature and chemical potential, which thereby enhanced the screening effect on the heavy quark bound states.
This leads to a decrease in the dissociation distance, making it easier for the quark-antiquark pairs to enter a deconfined state and transition into free quarks. 

\begin{figure*}[tbp!]
    \centering
    \includegraphics[width=1.0\textwidth]{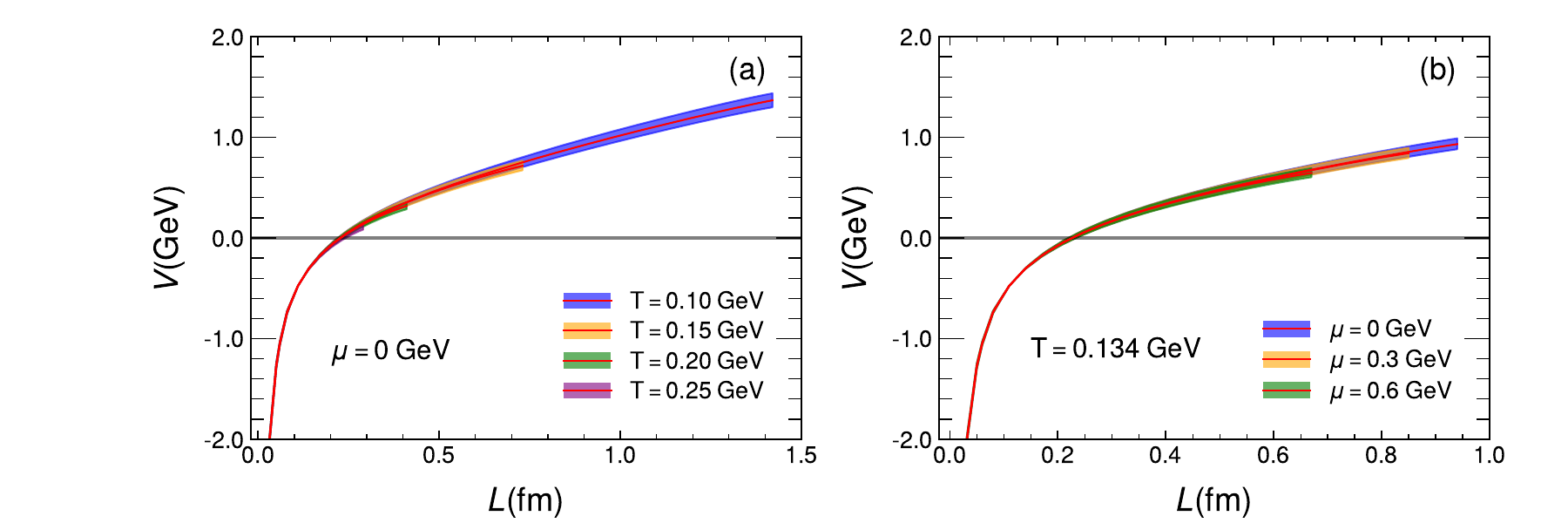}
    \caption{(a) The dependence of the potential energy $V$ of quark-antiquark pairs on the interquark distance $L$ at different temperatures when $\mu=0$. (b) The dependence of the potential energy $V$ of quark-antiquark pairs on the interquark distance $L$ at different chemical potential when $T=0.134\;\mathrm{GeV}$. The shaded area with color represents the 95\% CL, and the red solid line denotes the result of the MAP calculation.
}
\label{L-F}
\end{figure*}

\subsection{Potential energy}

In this subsection, we investigate the potential energy of heavy quark-antiquark pairs, a crucial physical quantity that characterizes the interaction between quark and antiquark. From recent pNRQCD\cite{Brambilla:2008cx} studies and one-loop hard-thermal-loop (HTL) calculations\cite{Laine:2006ns}, it's pointed out that the heavy quark potential at finite temperatures will also develop a nonvanishing imaginary part reflecting the Laudau damping and color singlet to octet transition. In this work we discuss only the real part of the potential and will leave the inclusion of imaginary part for future works. This real part potential energy typically consists of a short-range Coulomb potential and a long-range linear potential. At short distances, the Coulomb potential dominates, expressed as $F(r) \propto -\frac{\alpha}{r}$, where $\alpha$ is the effective coupling constant of the strong interaction, the Coulomb potential reflects the attractive force between the quark and antiquark, similar to the Coulomb potential in electromagnetic interactions, but mediated by the color charge of the strong interaction. At long distances, the potential energy gradually shifts to being dominated by the linear potential, reflecting the binding mechanism of quark pairs, expressed as $F(r)\propto\sigma r$, where $\sigma$ is the string tension constant. This linear potential represents the binding effect of the "string" formed by the gluon field between the quark and antiquark, corresponds to manifestation of confinement and is a significant feature of the strong interaction, ensuring that quarks and antiquarks remain bound together at large distances. 

\begin{figure}[h]
    \centering
    \includegraphics[width=0.45\textwidth]{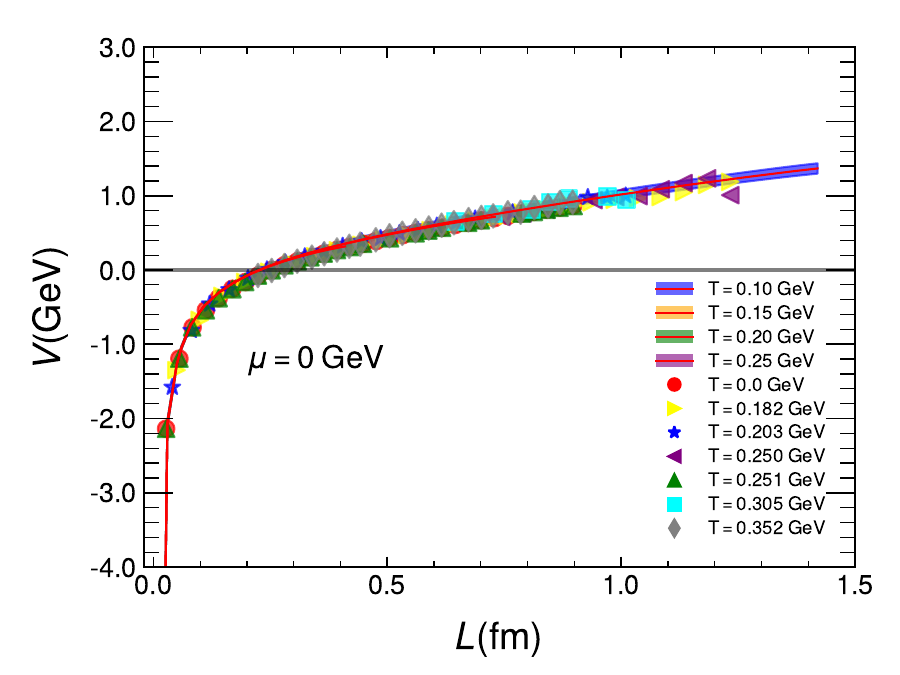}
    \caption{A comparison is made between the potential energy of quark-antiquark pairs from model calculation results and lattice data \cite{Bazavov:2023dci} for the case of \(N_f = 2 + 1\) and \(\mu = 0\). The lattice data is represented by various shapes of points. The shaded area with color represents the 95\% CL, and the red solid line denotes the result of the MAP calculation.
}
\label{V2}
\end{figure}

As shown in Fig. \ref{L-F}, when the separation distance between quarks is very small, the potential energy is primarily influenced by the Coulomb potential, resulting in a negative value. As the quark separation distance increases, the potential energy gradually transitions to being dominated by the linear potential. In Fig. \ref{L-F}, We calculated the results for the 95\% CL (shaded area with color) and the MAP (red solid line) setup, sub-figure (a) displays the calculation results of the effect of different temperatures on the potential energy with the chemical potential fixed at zero, while sub-figure (b) presents the calculation results of the effect of chemical potential on the potential energy with the temperature fixed at $T=0.134 \;\mathrm{GeV}$. Our calculations indicate that with the increase of system temperature and chemical potential, the linear potential component of the quark pair slightly decreases but does not show clear color screening effect. This phenomenon may be related to the tendency for quark pairs to dissociate due to the increased thermal energy, which reduces the length of the linear component of the potential energy. Meanwhile, the Coulomb potential component remains almost unaffected, indicating that the short-range strong interactions remain stable under higher temperatures and chemical potentials.
As shown in Fig. \ref{V2}, we compare the results from our model with the recent lattice data \cite{Bazavov:2023dci}, where the data points represent the lattice results, and the colored shaded regions along with the red lines depict our model's predictions, this comparison clearly depicts a significant resemblance between the Bayesian holographic model's results and the lattice data for heavy quark potential.

\begin{figure*}[tbp!]
    \centering
    \includegraphics[width=1.0\textwidth]{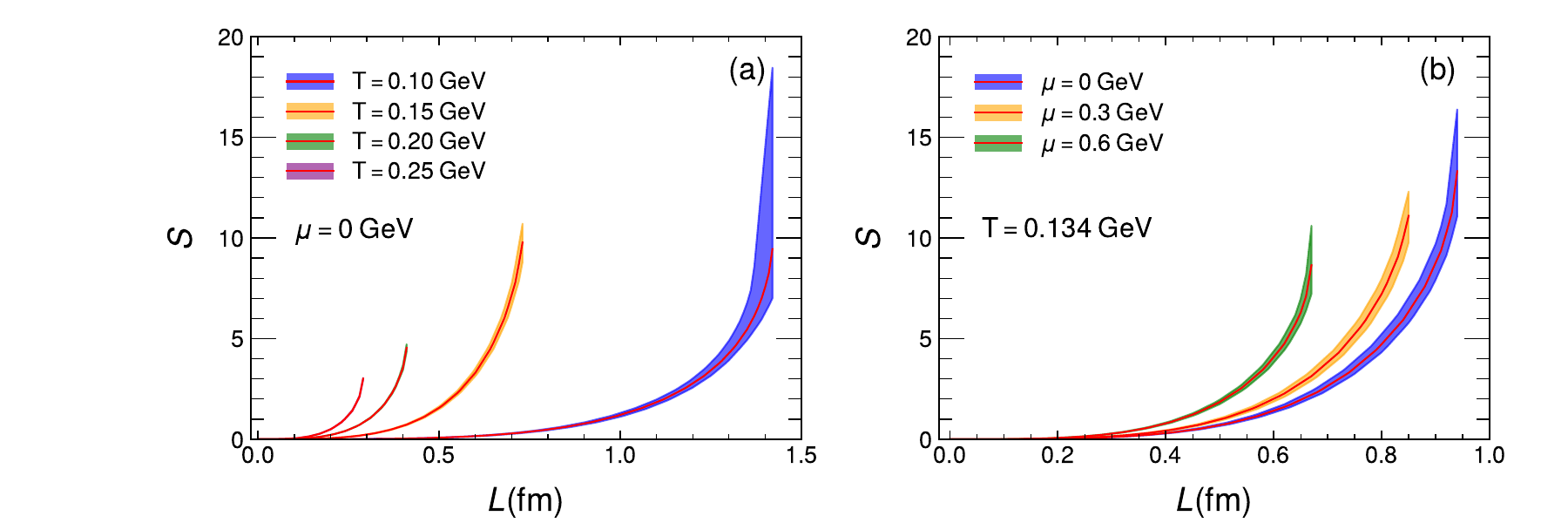}
    \caption{(a) The dependence of the entropy $S$ of quark-antiquark pairs on the interquark distance $L$ at different temperatures when $\mu=0$. (b) The dependence of the entropy $S$ of quark-antiquark pairs on the interquark distance $L$ at different chemical potential when $T=0.134\;\mathrm{GeV}$. The shaded area with color represents the 95\% CL, and the red solid line denotes the result of the MAP calculation.
}
\label{L-S}
\end{figure*}

\begin{figure*}[tbp!]
    \centering
    \includegraphics[width=1.0\textwidth]{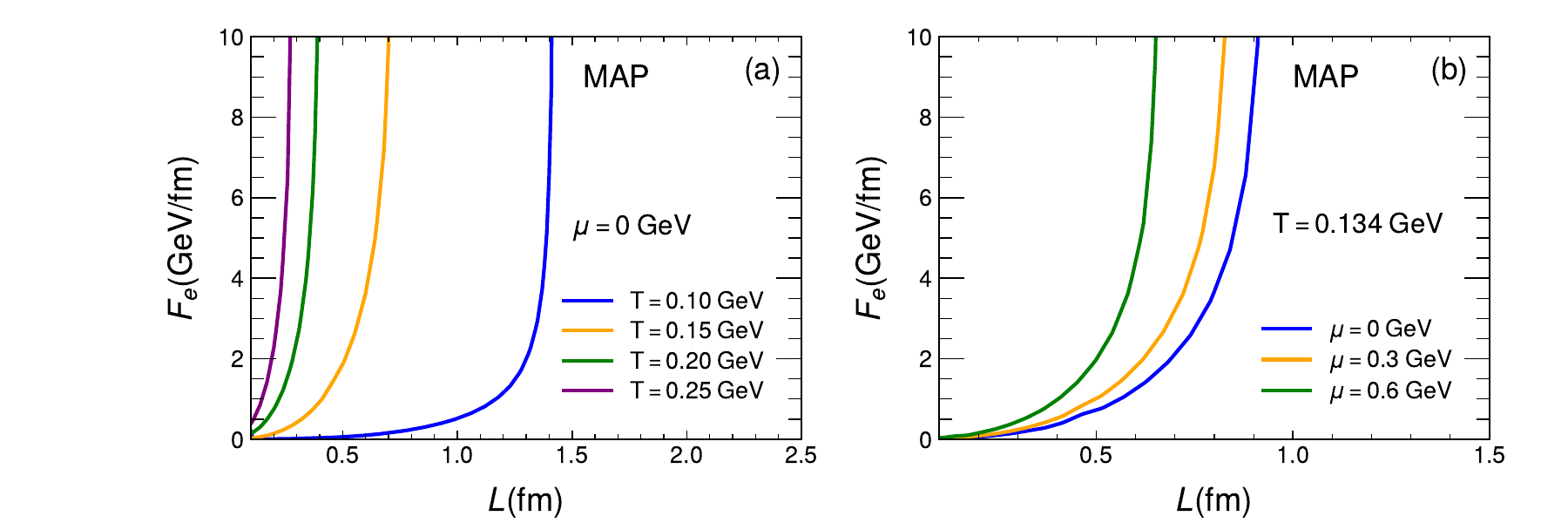}
    \caption{When selecting MAP value, (a) The dependence of the entropy force $F_{e}$ on interquark distance $L$ for the quark-antiquark pair at different temperatures when $\mu=0$. (b) The dependence of the entropy force $F_{e}$ on interquark distance $L$ for the quark-antiquark pair at different chemical potentials when $T=0.134\;\mathrm{GeV}$.
}
\label{L-Fe-MAP}
\end{figure*}

\subsection{Entropy}

\begin{figure*}[tbp!]
    \centering
    \includegraphics[width=1.0\textwidth]{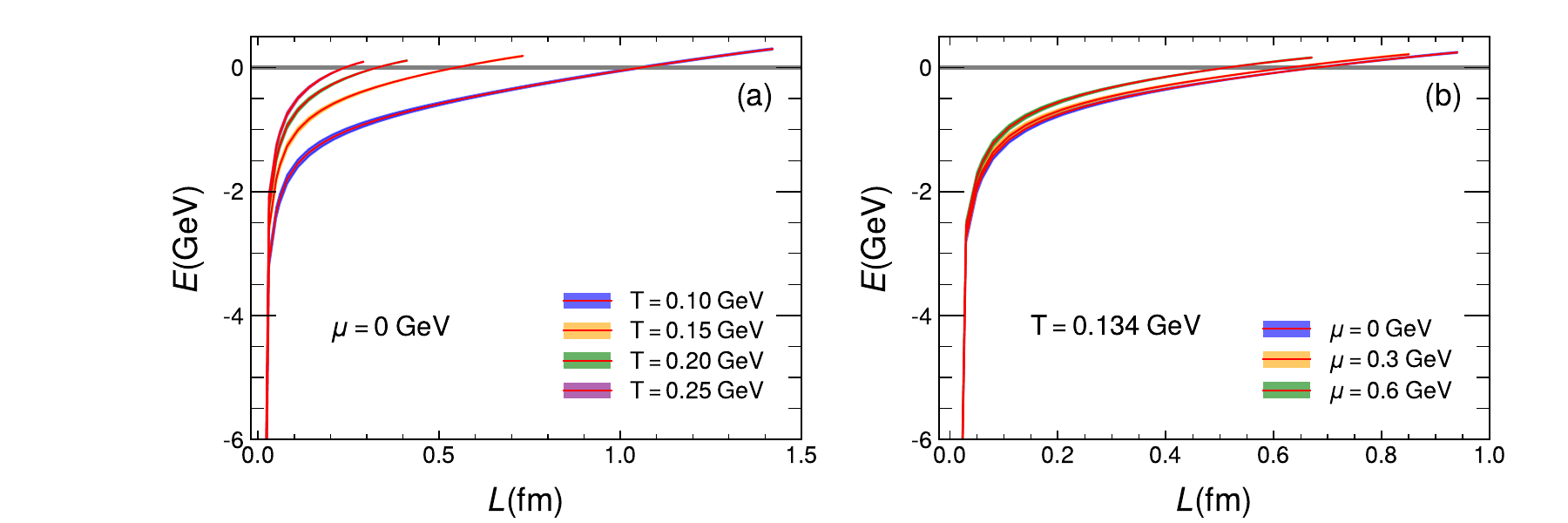}
    \caption{(a) The dependence of the binding energy $E$ of quark-antiquark pairs on the interquark distance $L$ at different temperatures when $\mu=0$. (b) The dependence of the binding energy $E$ of quark-antiquark pairs on the interquark distance $L$ at different chemical potential when $T=0.134\;\mathrm{GeV}$. The shaded area with color represents the 95\% CL, and the red solid line denotes the result of the MAP calculation.
}
\label{L-E}
\end{figure*}

\begin{figure*}[tbp!]
    \centering
    \includegraphics[width=1.0\textwidth]{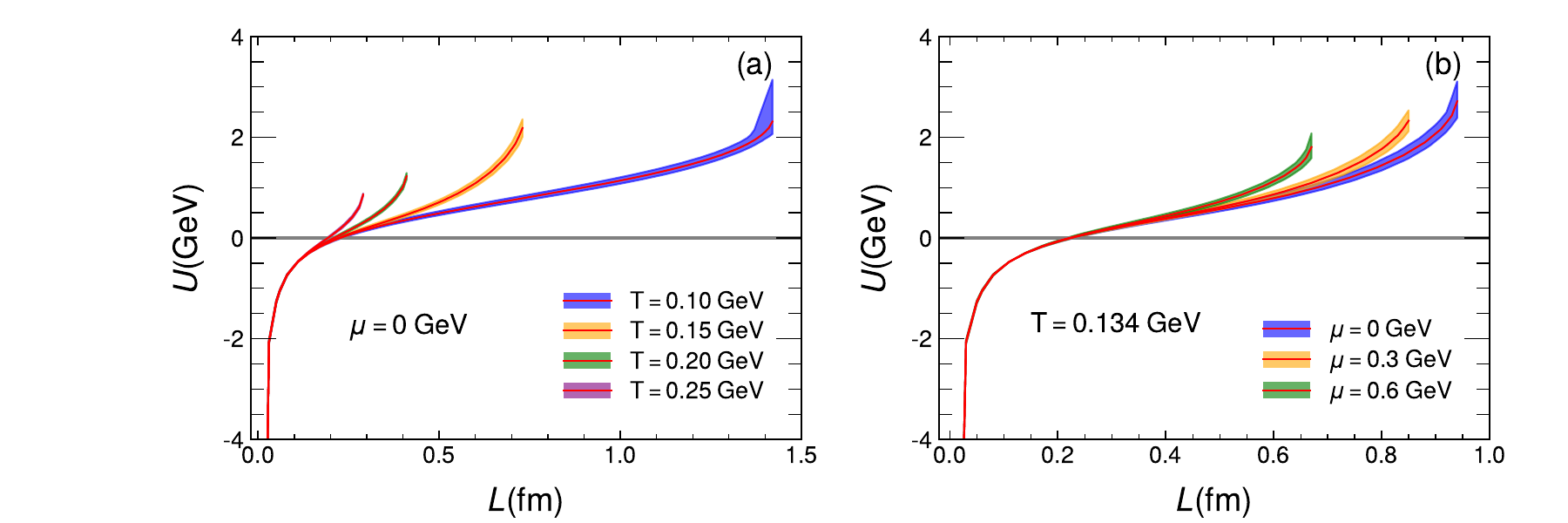}
    \caption{(a) The dependence of the internal energy $U$ of quark-antiquark pairs on the interquark distance $L$ at different temperatures when $\mu=0$. (b) The dependence of the internal energy $U$ of quark-antiquark pairs on the interquark distance $L$ at different chemical potential when $T=0.134\;\mathrm{GeV}$. The shaded area with color represents the 95\% CL, and the red solid line denotes the result of the MAP calculation.
}
\label{L-U}
\end{figure*}

In this subsection, we systematically investigate the entropy of heavy quark anti-quark pairs, an important physical quantity used to characterize the degree of disorder in the micro-states induced by the interactions between quarks and antiquarks. Entropy not only provides crucial insights into the thermodynamic properties of the heavy quark system but also reveals the underlying mechanisms of its dynamical behavior. The specific expression for the entropy is detailed in Eq. (\ref{eq30}).
Furthermore, we examine how the entropy varies with the interquark distance $L$ and analyze its behavior under different temperature and chemical potential conditions. The relevant results are presented in Fig. \ref{L-S}, We calculated the results for the 95\% CL (shaded area with color) and the MAP (red solid line), subfigure (a) displays the calculation results of the effect of different temperatures on the entropy with the chemical potential fixed at zero, while subfigure (b) presents the calculation results of the effect of chemical potential on the entropy with the temperature fixed at $T=0.134 \;\mathrm{GeV}$. Our research indicates that as the temperature and chemical potential increase, the value of the entropy also rises. This finding leads us to conclude that in environments with higher temperature and chemical potential, the production rate of heavy quark-antiquark pairs is significantly suppressed.

Under specific temperature or chemical potential conditions, an increase in the distance between quarks leads to a significant rise in entropy. This increase in entropy directly results in a larger entropy force, expressed as $F_{e}=T\frac{\partial S}{\partial L}$, as shown in Fig. \ref{L-Fe-MAP}. As the dissociation distance approaches the maximum value $L_{\mathrm{max}}$, the entropy  force tends to infinity. This phenomenon indicates that the increase in entropy is not merely a thermodynamic characteristic; it is closely related to the dynamical dissociation process of quark-antiquark pairs. As discussed in detail in reference \cite{Kharzeev:2014pha,Zhu:2024dwx,Wu:2022ufk}, the enormous entropy force is considered a key factor driving the dissociation of heavy quark-antiquark pairs, further emphasizing the importance of entropy in the study of quark physics.

\subsection{Binding energy}

In this section, we use the Eq. (\ref{eq31}) to calculate the binding energy of heavy quark-antiquark pairs as a function of the distance $L$ between quarks at different temperatures and chemical potentials. As shown in Fig. \ref{L-E}, We calculated the results for the 95\% CL (shaded area with color) and the MAP (red solid line), subfigure (a) displays the calculation results of the effect of different temperatures on the binding energy with the chemical potential fixed at zero, while Subfigure (b) presents the calculation results of the effect of chemical potential on the binding energy with the temperature fixed at $T=0.134\;\mathrm{GeV}$. The binding energy increases with rising temperature and chemical potential. When $L$ reaches the critical value $L_{c}$ (where $L_{c}\leq L_{\mathrm{max}}$) , the binding energy becomes zero, indicating that the potential energy of the bound heavy quark-antiquark pair is equal to the free energy of the unbound pair, i.e., $E(L_{c})=0$. This signifies that the heavy quarkonium undergoes a phase transition from a connected string to a disconnected string. When $L$ exceeds this critical value $L_{c}$, the binding energy becomes positive, meaning that the potential energy of the bound heavy quark-antiquark pair is higher than the free energy of the unbound pair. Moreover, the increase in temperature and chemical potential not only enhances the binding energy but also reduces $L_{c}$, indicating that the rise in temperature and chemical potential promotes the dissociation of heavy quarkonium.

\subsection{Internal energy}

In this section, we employ Eq. (\ref{eq33}) to calculate the relationship between the internal energy of heavy quark-antiquark pairs and the interquark distance $L$ under varying temperatures and chemical potentials.
As depicted in Fig. \ref{L-U}, We calculated the results for the 95\% CL (shaded area with color) and the MAP (red solid line), subfigure (a) displays the calculation results of the effect of different temperatures on the internal energy with the chemical potential fixed at zero, while subfigure (b) presents the calculation results of the effect of chemical potential on the internal energy with the temperature fixed at $T=0.134 \;\mathrm{GeV}$. The results indicate that at smaller interquark distances $L$, the internal energy $U$ is negative and remains relatively unaffected by changes in temperature and chemical potential. However, at larger interquark distances $L$, the internal energy becomes positive and increases with rising temperature and chemical potential. According to the definition of internal energy, this behavior suggests that at short interquark distances, the internal energy is predominantly governed by potential energy, whereas at larger distances, entropy contributions play a significant role.

\section{THERMODYNAMICS OF SINGLE QUARK}
\label{sec4}

In this section, we will  investigate the free energy of single quark as shown in Eq. (\ref{eq32}), as well as the entropy and internal energy, with the expressions as follows:

\begin{align}
S_{Q}=-\frac{\partial F_{Q}}{\partial T}=-\frac{\partial F_{Q}}{\partial z_{0}}\frac{\partial z_{0}}{\partial T}
\end{align}
\begin{align}
U_{Q}=F_{Q}+TS_{Q}
\end{align}

\begin{figure}[h]
    \centering
    \includegraphics[width=0.45\textwidth]{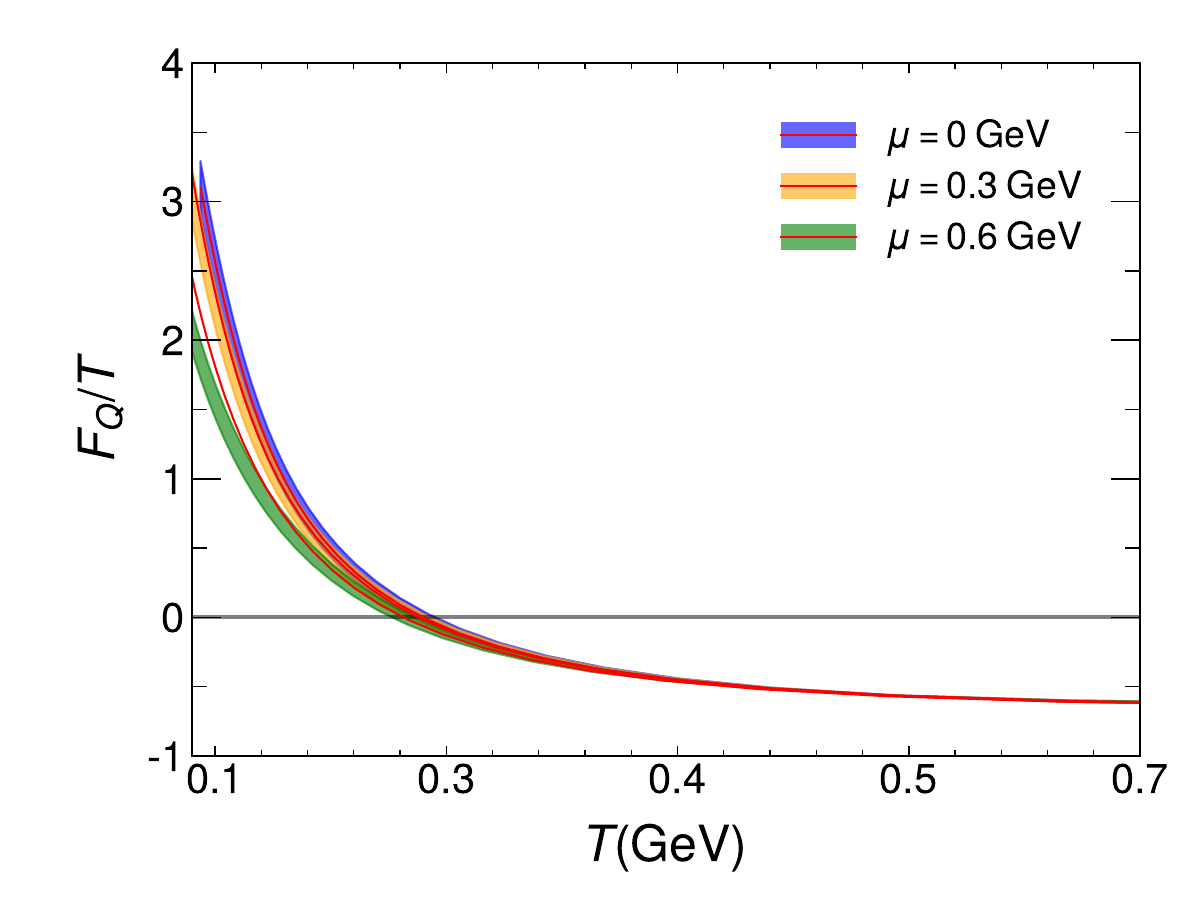}
    \caption{The dependence of the single quark free energy on the temperature for different chemical potentials. The shaded area with color represents the 95\% CL, and the red solid line denotes the result of the MAP calculation.
}
\label{Singlequark-F}
\end{figure}

\begin{figure}[h]
    \centering
    \includegraphics[width=0.45\textwidth]{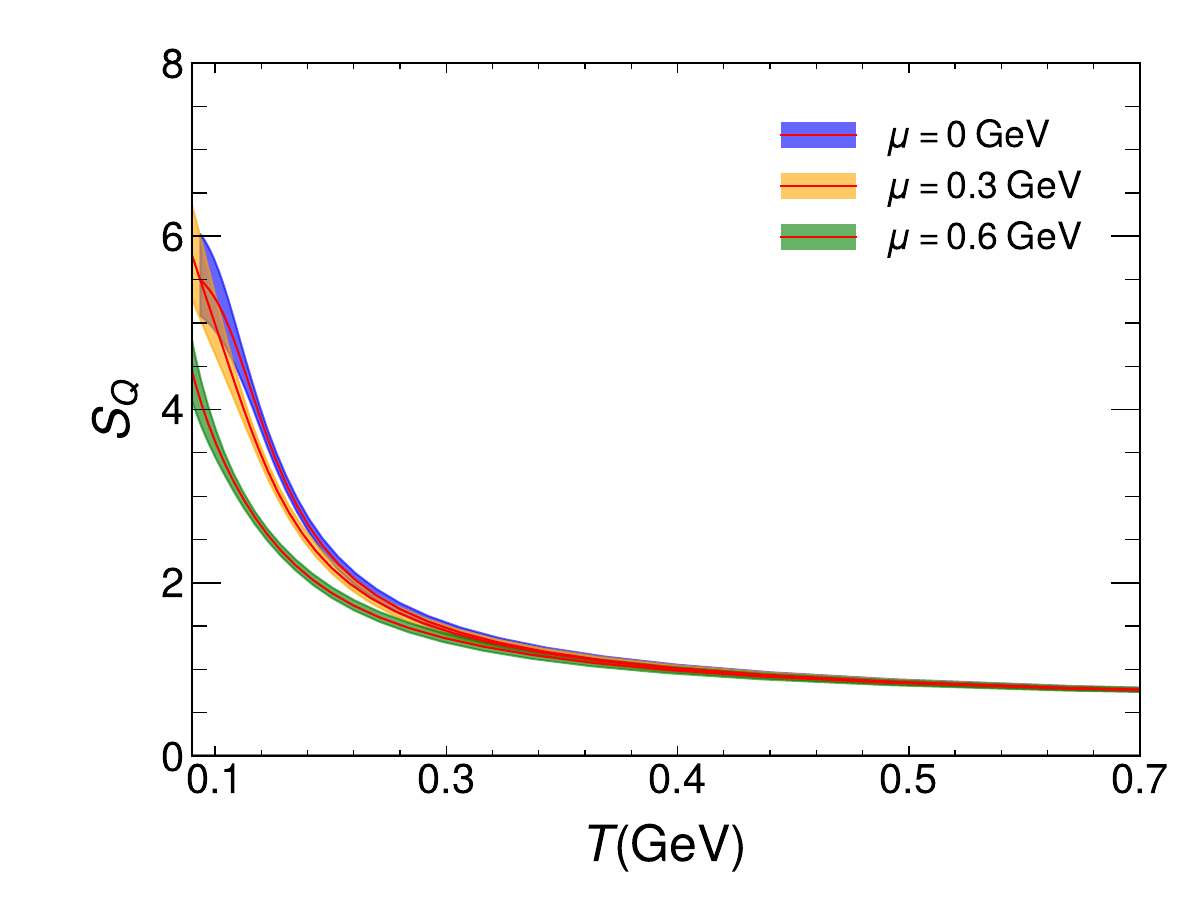}
    \caption{The dependence of the single quark entropy on the temperature  for different chemical potentials. The shaded area with color represents the 95\% CL, and the red solid line denotes the result of the MAP calculation.
}
\label{Singlequark-S}
\end{figure}

\begin{figure}[h]
    \centering
    \includegraphics[width=0.45\textwidth]{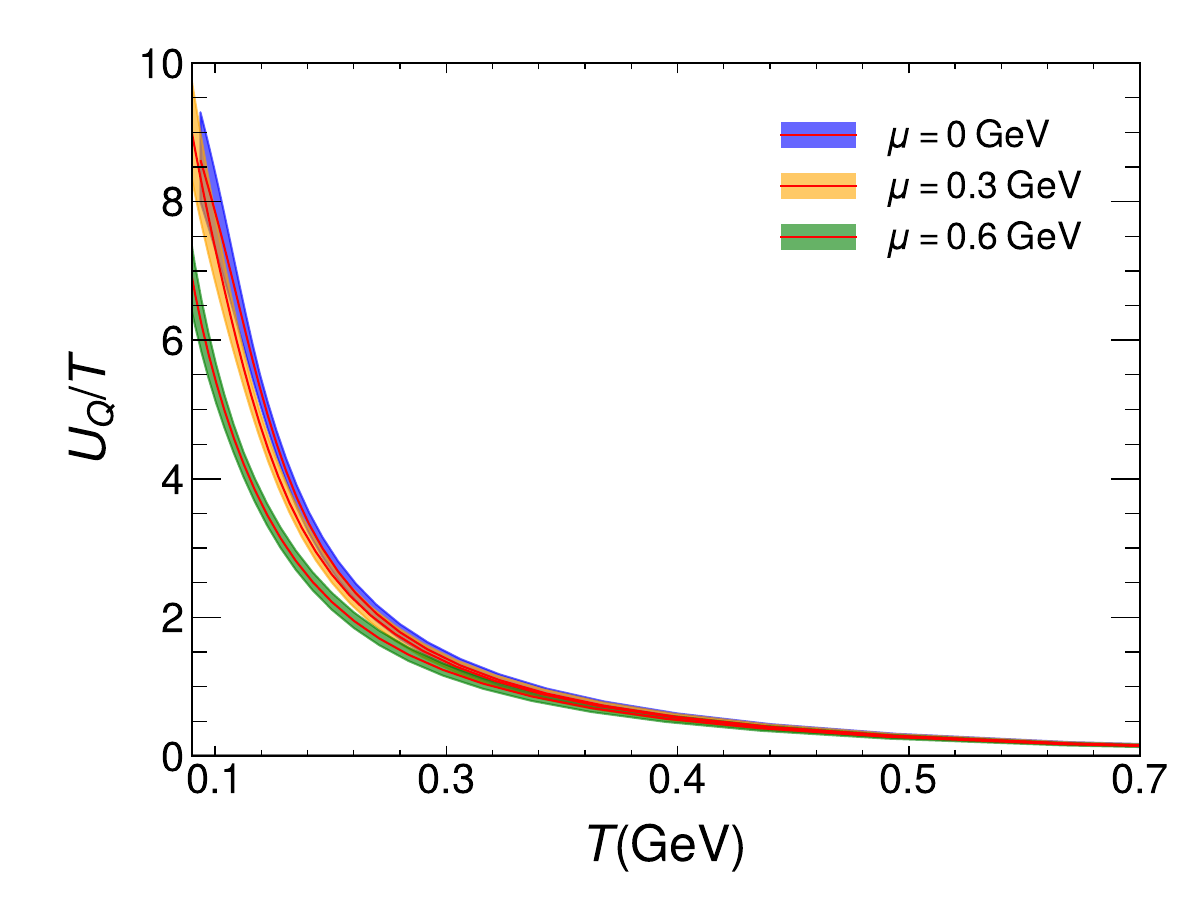}
    \caption{The dependence of the single quark internal energy on the temperature for different chemical potentials. The shaded area with color represents the 95\% CL, and the red solid line denotes the result of the MAP calculation.
}
\label{Singlequark-U}
\end{figure}

In this section, we will investigate the effect of chemical potential on the thermodynamic quantities of a single quark. We have calculated the results for the 95\% CL (shaded area with color) and the maximum a posteriori estimate (red solid line). The free energy of a single quark is shown in Fig. \ref{Singlequark-F}, where it can be seen that an increase in chemical potential leads to an increase in free energy, and 
$F_{Q}/T$ will approach a conformal situation in the high-temperature limit. The entropy of a single quark is shown in Fig. \ref{Singlequark-S}, where the results indicate that $S_{Q}$ also increases with the chemical potential and will tend toward conformal limit  in the high-temperature limit. The internal energy of a single quark is shown in Fig. \ref{Singlequark-U}, and the results also indicate that the internal energy increases with the chemical potential, with $U_{Q}/T$ also approaching the conformality  limit in the high-temperature limit.

\section{SUMMARY}
\label{sec5}

In this paper, we investigate the effects of temperature and chemical potential on the dissociation distance, potential energy, entropy, binding energy, and internal energy of heavy quarkonium based on a Bayesian holographic QCD model. Our study finds that an increase in temperature and chemical potential decrease the dissociation distance and suppress the heavy quark potential. When the interquark distance $L$ is small, the effects of temperature and chemical potential on the potential energy are minimal, but they become more pronounced when the interquark distance $L$ is large. Increasing temperature and chemical potential result in an increase in entropy, Within the dissociation distance, the entropy also increases significantly with the interquark distance, leading to a larger entropic force. This larger entropy force facilitates the dissociation of heavy quark-antiquark bound states. Correspondingly, the binding energy reaches zero at smaller interquark distance $L$ as temperature and chemical potential increase, indicating that under high temperature and high chemical potential, the binding force of heavy quark-antiquark pairs becomes weaker. The internal energy increases with rising temperature and chemical potential, at smaller interquark distance $L$, the internal energy is primarily dominated by potential energy, while at larger interquark distance $L$, it is dominated by $TS$. The effects of temperature and chemical potential on the internal energy are minimal at small quark interquark distance, but become significant at larger interquark distance.

Our results robustly demonstrate that elevated temperature and chemical potential accelerate heavy quarkonium dissociation by enhancing entropic forces and suppressing binding energy. These findings deepen the understanding of quark-gluon plasma (QGP) signatures in heavy-ion collisions and provide a framework for probing QCD matter under extreme conditions. The consistency of single-quark free energy and entropy trends further supports the generality of our conclusions.

While our model captures essential features of quarkonium dissociation, several limitations warrant discussion: The Bayesian framework relies on specific assumptions about the holographic dual, which may not fully capture non-perturbative QCD effects; The chemical potential range studied here is applicable to heavy-ion collision scenarios, but extensions to neutron star matter or other high-density systems require further exploration.

\section{Acknowledgements
\label{sec:acknowledgements}
}

This work is supported in part by the National Key Research and
 Development Program of China under Contract No. 2022YFA1604900.   This work is also supported in part by the National Natural Science Foundation of China (NSFC) Grant Nos: 12405154, 12235016, 12221005,  12435009, 12275104.and the Strategic Priority Research Program of Chinese Academy of Sciences under Grant No XDB34030000, the Fundamental Research Funds for the Central Universities, Open fund for Key Laboratories of the Ministry of Education under Grants No.QLPL2024P01, CUHK-Shenzhen university development fund under grant No.\ UDF01003041 and UDF03003041, Shenzhen Peacock fund under No.\ 2023TC0007, and the European Union -- Next Generation EU through the research grant number P2022Z4P4B ``SOPHYA - Sustainable Optimised PHYsics Algorithms: fundamental physics to build an advanced society'' under the program PRIN 2022 PNRR of the Italian Ministero dell'Universit\`a e Ricerca (MUR).

\bibliographystyle{apsrev4-2}
\bibliography{clv3}

\end{document}